\documentclass[a4paper,11pt]{article}
\usepackage{pos}
\title{Energy-momentum tensor in the 2d $O(3)$ non-linear sigma model on the lattice}
\author*[a]{Mika Lauk}
\author[a,b]{Agostino Patella}

\affiliation[a]{Humboldt Universit\"at zu Berlin, Institut f\"ur Physik, \\ Zum Grossen Windkanal 6, 12489 Berlin, Germany}
\affiliation[b]{DESY, Platanenallee 6, D-15738 Zeuthen, Germany}

\emailAdd{mika.akim.lauk@hu-berlin.de}
\abstract{
The long-term goal of this project is the non-perturbative renormalization of the energy-momentum tensor in the 2d $O(3)$ non-linear sigma model using different methods which have been developed for QCD applications.
As a first step, we have identified all operators that mix with the energy-momentum tensor once a lattice discretization is employed, that is all which are compatible with power counting and with the symmetries of the theory. Since these operators are constrained by non-linear Ward identities arising from the non-linear realization of the $O(3)$ symmetry, this is not entirely straightforward on the technical level. We have also outlined the basics of ongoing numerical simulations with shifted boundary conditions and an optimized constraint action to minimize lattice artifacts.
}

\FullConference{The 41st International Symposium on Lattice Field Theory (LATTICE2024)\\
 28 July - 3 August 2024\\
Liverpool, UK\\}

\begin{document}
\maketitle
\section{Introduction}
Two dimensional $O(N)$ non-linear sigma models are an interesting class of quantum field theories that are exactly solvable \cite{zamolodchikov_factorized_1979}. For $N=2$ they are conformal and for $N>3$ they display asymptotic freedom. These properties makes them interesting toy models for QCD. Additionally, they can be extended supersymmetrically, providing a playground to investigate aspects of strings in $AdS$ \cite{Costa:2022ezw}, whose worldsheet discretization remains a challenge \cite{Bliard:2022kne}. 

One possible starting point of interest in dealing with these models is the renormalization of the energy-momentum tensor in perturbation theory and non-perturbatively by means of lattice simulations. The energy-momentum tensor is an interesting observable in quantum field theory, as it is not only the Noether current associated to spacetime translations, but currents associated with other space-time symmetries can be related to it. It carries important information about the theory such as the equation of state and the conformal anomaly.

For now, we will focus on the case $N=3$, although it has to be stressed that the results in these proceedings were obtained keeping $N$ quite general and can be easily extended. In section 2 we will shortly describe what the $O(3)$ non-linear sigma model is and how its renormalization works, before moving on to applying this in section 3 in the context of the energy-momentum tensor. We will concisely outline some features of ongoing numerical simulations in section 4.

Throughout this work the Einstein convention is only used for indices with a superscript, not a subscript.

\section{Sigma models and their Renormalization}

The $O(3)$ non-linear sigma model is defined through a scalar field $\phi$ mapping a flat two-dimensional spacetime into the target space $S^{2}$. The discretized Euclidean action then takes the simple form
\begin{equation}
    S = \frac{1}{2 g_0^2} \sum_{x,\mu} a^2 \; \partial^f_\mu \phi^I(x) \partial^f_\mu \phi^I(x)
\end{equation}
where $\partial^f$ denotes the discretized forward derivative and the field satisfies the constraint $\phi^I \phi^I = 1$. Expectation values of observables are defined by means of the path integral formula
\begin{gather}
    \langle P \rangle = \frac{1}{Z} \int \left[ \prod_x d^3\phi(x) \, \delta(\phi^I(x) \phi^I(x) - 1 ) \right] e^{-S(\phi)} P(\phi) ,
\end{gather}
where $d^3\phi(x) \, \delta(\phi^I(x) \phi^I(x) - 1 )$ is nothing but the canonical measure on the sphere $S^{2}$ up to a normalization factor. Both the action and the integration measure are invariant under an $O(3)$ symmetry which acts by rotating the components of the field. Ultraviolet divergences in $n$-point functions of the field are removed by renormalizing the field and the coupling constant as in
\begin{equation}
    \phi_r = Z^{-1/2} \phi ,
    \qquad
    g_r^2 = Z_1^{-1} g_0^2 .
\end{equation}
This statement is proven in \cite{ZJB} in dimensional regularization at all orders in the perturbative expansion. Zinn-Justin outlines in his book \cite{ZJbook} how the same renormalization pattern applies to the lattice regularization.

We are interested in the renormalization of $O(3)$-invariant composite operators in this theory. If the target renormalized operator $\mathcal{O}_r(x)$ has mass dimension $D$, then it is generally written as a linear combination of bare operators $\mathcal{O}(x)$ with mass dimension $D_\alpha \le D$, i.e.
\begin{equation}\label{eq:Or}
    \mathcal{O}_r(x)=\sum_{D_\alpha \leq D} Z^\alpha \mathcal{O}^\alpha(x).
\end{equation}
The operators $\mathcal{O}^\alpha(x)$ are all those that are compatible with the symmetries of $\mathcal{O}_r$, keeping in mind that only the symmetries of the regularized theory matter. What this statement exactly means can be understood at all orders in the perturbative expansion. One can conjecture that, since the theory is asymptotically free and the renormalization properties are dictated by the large-energy regime of the theory, the renormalization pattern found by means of a perturbative analysis should hold also at the non-perturbative level. Thus one should first formulate a perturbative expansion and carry out an all orders renormalization thereof to find the general form of the renormalized operator, before calculating the renormalization constants non-perturbatively. At the perturbative level, the symmetry analysis is complicated by the fact that one needs to choose a parametrization of the sphere $S^2$, e.g. by setting
\begin{equation}
    \phi = ( \boldsymbol{\pi}, \sigma ) , \quad \text{where} \quad \boldsymbol{\pi} = ( \pi^1,\pi^{2} ), \ 
    \sigma = \sqrt{1 - \pi^i \pi^i}.
\end{equation}
In this parametrization, the direction $e_3=(0,0,1)$ in field space plays a special role. The $O(2)$ subgroup of $O(3)$ which leaves $e_3$ invariant acts linearly on the field $\boldsymbol{\pi}$ and leaves $\sigma$ invariant. Invariance under $O(2)$ implies that only $O(2)$-invariant operators can appear in the right-hand side of \eqref{eq:Or}. The remaining symmetries act non-linearly on $\boldsymbol{\pi}$. Given small parameters $\omega = (\omega^1,\omega^2)$, the rotations on the plane defined by $e_3$ and $(\omega^1,\omega^2,0)$ act infinitesimally as
\begin{equation}
    \delta_\omega \pi^i=\omega^i \sigma \qquad \delta_\omega\sigma=-\omega^i \pi^i .
\end{equation}
The associated Ward identities can be expressed as quadratic first-order differential equations for the generating functional $\Gamma$ of 1PI diagrams, coupled to sources for the operators $\sigma$ and $O^\alpha$. In order to understand the renormalization of $O^\alpha$, one needs to understand the structure of the divergent part of $\Gamma$ which is constrained in a non-trivial way by the Ward identities. One finds that the operators appear in the right-hand side of \eqref{eq:Or} are general local functions of $\phi$ and
\begin{gather}
    \eta = \frac{\hat{\Box} \sigma}{\sigma} ,
\end{gather}
with the property that they are invariant under $O(3)$ rotations of $\phi$ at fixed $\eta$.

\section{Renormalization of the energy-momentum tensor}
The energy-momentum tensor can be defined as the Noether current obtained from the invariance of the theory under translations. In the continuum, it is given by
\begin{equation}
    T_{\mu\nu}(x) = \partial_{\mu} \phi^I(x) \partial_{\nu} \phi^I(x) - \frac{1}{2} \partial_{\eta} \phi^I(x) \partial_{\eta} \phi^I(x).
\end{equation}
When a lattice discretization is considered, translational symmetry is broken and this naive expression can no longer be used. Keeping in mind that we do not want to give up locality of the theory, we shall not aim for an exact symmetry on the lattice, but to recover it in the continuum limit  \cite{caracciolo_energy-momentum_1988}. In other words,
we want to obtain the right lattice energy-momentum tensor, and properly renormalize it to achieve this goal.

In the continuum, the energy-momentum tensor decomposes into two irreducible representations of the $O(2)$ group of Euclidean spacetime rotations: the trace and the traceless symmetric parts. On the lattice this symmetry is broken down to its $D_4$ subgroup of reflections and $90^\circ$ rotations. As a consequence, the traceless symmetric representation branches off into two irreducible representation of $D_4$: the traceless diagonal and off-diagonal parts. These parts will necessarily renormalize independently. Therefore, the renormalized energy-momentum tensor can be written as
\begin{equation}
     \left[T_{\mu\nu} (x) \right]_r = \sum_{\alpha = 1}^{4} Z^\alpha \left\{T_{\mu\nu}^\alpha (x) - \langle T_{\mu\nu}^\alpha (x) \rangle \right\},
\end{equation}
where
\begin{equation}
    \begin{aligned}
        T^1_{\mu\nu}(x) &= \delta_{\mu\nu} \sum_\rho \partial^s_\rho \phi^I(x) \partial^s_\rho \phi^I(x) , \\
        T^2_{\mu\nu}(x) &= \delta_{\mu\nu} \left[ \partial^s_\mu \phi^I(x) \partial^s_\nu \phi^I(x) - \frac{1}{2} \sum_\rho \partial^s_\rho \phi^I(x) \partial^s_\rho \phi^I(x)\right] , \\
        T^3_{\mu\nu}(x) &= (1 - \delta_{\mu\nu}) \partial^s_\mu \phi^I(x) \partial^s_\nu \phi^I(x) , \\
        T^4_{\mu\nu}(x) &= \frac{\delta_{\mu\nu}}{\sigma(x)} \hat{\Box} \sigma(x),
        \end{aligned}
\end{equation}
$\partial^s$ being the discretized symmetric derivative and $\hat{\Box}$ the laplacian.
Naively, using counting of canonical dimensions, the only other operators that might mix with the energy-momentum tensor are $\delta_{\mu\nu} \phi \hat{\Box} \phi$ and $\phi \partial^s_\mu \partial^s_\nu \phi$. However, in the continnum, these operators are linear combination of the ones given above and we do not need to consider them. For instance
\begin{equation}
        0 = \partial_{\mu} (\phi^I \phi^I) = 2 \phi^I \partial_{\mu} \phi^I
\end{equation}
and thus 
\begin{equation}
        0 = \sum_\mu \partial_{\mu} (\phi^I \partial_{\mu} \phi^I) = \sum_\mu \partial_{\mu} \phi^I \partial_{\mu} \phi^I + \phi^I \Box \phi^I.
\end{equation}

We are interested in correlation functions of the form
\begin{equation}
    \label{emtcorr}
    \left\langle \mathcal{P} \left[T_{\mu\nu} (0) \right]_r \right\rangle = \sum_{\alpha = 1}^{4} Z^\alpha \left\langle \mathcal{P} \left\{T_{\mu\nu}^\alpha (0) - \langle T_{\mu\nu}^\alpha (0) \rangle \right\} \right\rangle,
\end{equation}
where $\mathcal{P}$ is a $O(3)$-invariant product of local operators at spacetime points away from the origin (in order to avoid contact terms with the energy-momentum tensor). This expression has the undesirable feature that the last term is not manifestly $O(3)$ invariant and depends on the choice of parametrization. However, in $O(3)$ invariant expectation values, we expect that we should be able to exchange insertions of $O(3)$ non-invariant operators with invariant ones. Thus, there should be an equivalent expression involving only invariant operators. Indeed we find such an expression by considering the Dyson-Schwinger equations for the polynomially bounded functional
\begin{equation}
    \pi^i(0) \mathcal{P}
\end{equation}
for each $i \in \{1,2\}$, taking the sum afterwards. We obtain the relation
\begin{equation}
\left\langle \pi_i(0) \frac{\delta \mathcal{P}}{\delta \pi_i(0)} \right\rangle + \frac{2}{a^2} \left\langle  \mathcal{P} \right\rangle =
\left\langle \mathcal{P} \frac{\hat{\Box} \sigma(0)}{\sigma(0)}  \right\rangle
- \left\langle \mathcal{P} \phi^I(0) \hat{\Box} \phi^I(0) \right\rangle
- \frac{1}{a^2} \left\langle \mathcal{P} \frac{\pi^i(0)\pi^i(0)}{\sigma^2(0)} \right\rangle.
\end{equation}
The first term vanishes under the assumption that $\mathcal{P}$ is a product of local operators at spacetime points away from the origin. The last term can be further rewritten using the Ward identity $\left\langle \delta_\omega \left(\frac{\pi_i}{\sigma} \mathcal{P} \right) \right\rangle = 0 $, which is equivalent to
\begin{equation}
    0 = \left\langle \mathcal{P} \right\rangle \delta^{ij} + \left\langle \mathcal{P} \frac{\pi^i(0)\pi^j(0)}{\sigma^2(0)} \right\rangle.
\end{equation}
From this, it follows that:
\begin{equation}
        \left\langle \mathcal{P} \frac{\hat{\Box} \sigma(0)}{\sigma(0)} \right\rangle = 
        \left\langle \mathcal{P} \phi^I(0) \hat{\Box} \phi^I(0) \right\rangle .
\end{equation}
Thus, we can write equation \eqref{emtcorr} equivalently as
\begin{equation}
    \begin{aligned}
        \left\langle \mathcal{P} \left[T_{\mu\nu} (0) \right]_r \right\rangle &= \sum_{\alpha = 1}^{3} Z^\alpha \left\langle \mathcal{P} \left\{T_{\mu\nu}^\alpha (0) - \langle T_{\mu\nu}^\alpha (0) \rangle \right\} \right\rangle \\
        &+ Z^4 \left\langle \mathcal{P} \left\{\phi^I(0) \hat{\Box} \phi^I(0) - \langle \phi^I(0) \hat{\Box} \phi^I(0) \rangle \right\} \right\rangle
    \end{aligned}
\end{equation}
which is manifestly $O(3)$ invariant. Moreover, as we have already noted, the operator $\phi \Box \phi$ is equivalent to $T^1$ in the continuum. Thus, its insertion can be absorbed by a redefinition of the coefficient $Z^1$. We conclude that, as long as it is inserted into correlation functions of arbitrary $O(3)$ invariant operators at pairwise-distinct points, the renormalized energy-momentum tensor takes the form:
\begin{equation}
    \left[\widetilde{T}_{\mu\nu} (x) \right]_r = \sum_{\alpha = 1}^{3} \widetilde{Z}^\alpha \left\{T_{\mu\nu}^\alpha (x) - \langle T_{\mu\nu}^\alpha (x) \rangle \right\},
\end{equation}
which we will use from now on.

\section{Numerical simulations}
We have found all operators that mix in renormalizing the energy-momentum tensor and outline a strategy for computing the renormalization constants non-perturbatively, as functions of $g_0^2$ \cite{symanzik_continuum_1983}. We use an optimized constraint action, a mix of the topological action and the standard lattice action. This action has the same quantum continuum limit and is shown to reduce large cutoff effects in the $O(3)$ model \cite{balog_puzzle_2010} \cite{balog_drastic_2012}. It is given by
\begin{equation}
    \begin{aligned}
        S_{\text {con}}[\phi]=\sum_{x, \mu}
        \begin{cases}
        \frac{1}{g_0^2}\left(1-\phi(x)^I \phi(x+\hat{\mu})^I\right)
        \quad
        & \phi(x)^I \phi(x+\hat{\mu})^I > -0.345\\
        \infty & \text{else.}
        \end{cases}
    \end{aligned}
\end{equation}
We perform the Monte-Carlo simulation using the Wolff cluster algorithm \cite{wolff_collective_1989}, reducing critical slowing down significantly. 

There are multiple ways to calculate the renormalization constants non-perturbatively. One that we are exploring is the use of shifted boundary conditions, which were successfully employed previously in the renormalization of the energy-momentum tensor in lattice Yang-Mills theory \cite{giusti_energy-momentum_2015}. Here, one formulates the theory at finite temperature in a moving frame, at Euclidean velocity $\xi$ \cite{giusti_implications_2013}. The partition function $Z$ has a path integral expression in terms of shifted boundary conditions in Euclidean time with extent $L_0$, given by
\begin{equation}
    \phi(L_0, x_1) = \phi(0, x_1 - \xi L_0).
\end{equation}
The corresponding free energy density in the infinite volume for this partition function is given by
\begin{equation}
    f(L_0,\xi) = \lim_{L_1 \rightarrow \infty} \frac{1}{L_0 L_1} \ln(Z(L_0,L_1,\xi))
\end{equation}
and fulfills the relation
\begin{equation}
    f(L_0,\xi) = f\left(L_0\sqrt{1+\xi^2},0\right).
\end{equation}
Ward Identities are obtained by taking derivatives with respect to the parameters $L_i$ and $\xi$, from which one can find relations between different components of the energy-momentum tensor. These can be used to calculate the renormalization constants and provide the advantage of this method. Namely, using them we only have to evaluate one-point functions of the energy-momentum tensor and can therefore also use an improved estimator for the two-point function of fields \cite{wolff_asymptotic_1990}.

\section{Conclusions \& Outlook}
We have found all operators mixing with the energy-momentum tensor in the 2d $O(3)$ non-linear sigma model in perturbation theory at all orders, finding an additional nontrivial operator that mixes with the scalar part that is not manifestly $O(3)$ invariant. However, employing Dyson-Schwinger equations and Ward identities from the $O(3)$ symmetry, we found that with respect to an insertion into correlation functions of arbitrary $O(3)$ invariant operators at pairwise-distinct points, it is equivalent to an insertion of an $O(3)$ invariant operator. Furthermore, this insertion can be absorbed by a redefinition of the renormalization constant of the scalar part.
We have outlined the ongoing effort of calculating the renormalization constants non-perturbatively using numerical simulations, using an optimized constraint action as well as shifted boundary conditions.

In the future, after finishing the calculation with shifted boundary conditions, we plan to perform a detailed comparison of this method to the use of the gradient flow and small flowtime expansion \cite{makino_renormalizability_2015}. We want to compare with integrability predictions (e.g. for the two-point function of the energy-momentum tensor), as well as investigate the trace anomaly and calculate the central charge in the $O(2)$ case. In the end, we also plan to generalize the program to supersymmetric target spaces.

\acknowledgments
The research of M.L. is funded by the Deutsche Forschungsgemeinschaft (DFG, German Research
Foundation) - Projektnummer 417533893/GRK2575 ”Rethinking Quantum Field Theory”.

\newpage
\bibliographystyle{JHEP}
\bibliography{main.bib}

\providecommand{\href}[2]{#2}\begingroup\raggedright\begin{thebibliography}{10}

\bibitem{zamolodchikov_factorized_1979}
A.B.~Zamolodchikov and A.B.~Zamolodchikov, \emph{{Factorized s Matrices in Two-Dimensions as the Exact Solutions of Certain Relativistic Quantum Field Models}}, \href{https://doi.org/10.1016/0003-4916(79)90391-9}{\emph{Annals Phys.} {\bfseries 120} (1979) 253}.

\bibitem{Costa:2022ezw}
I.~Costa, V.~Forini, B.~Hoare, T.~Meier, A.~Patella and J.H.~Weber, \emph{{Supersphere non-linear sigma model on the lattice}}, \href{https://doi.org/10.22323/1.430.0367}{\emph{PoS} {\bfseries LATTICE2022} (2023) 367} [\href{https://arxiv.org/abs/2212.11586}{{\ttfamily 2212.11586}}].

\bibitem{Bliard:2022kne}
G.~Bliard, I.~Costa, V.~Forini and A.~Patella, \emph{{Lattice perturbation theory for the null cusp string}}, \href{https://doi.org/10.1103/PhysRevD.105.074507}{\emph{Phys. Rev. D} {\bfseries 105} (2022) 074507} [\href{https://arxiv.org/abs/2201.04104}{{\ttfamily 2201.04104}}].

\bibitem{ZJB}
E.~Br\'ezin, J.~Zinn-Justin and J.C.~Le~Guillou, \emph{Renormalization of the nonlinear $\ensuremath{\sigma}$ model in $2+\ensuremath{\epsilon}$ dimensions}, \href{https://doi.org/10.1103/PhysRevD.14.2615}{\emph{Phys. Rev. D} {\bfseries 14} (1976) 2615}.

\bibitem{ZJbook}
J.~Zinn-Justin, \emph{Quantum Field Theory and Critical Phenomena: Fifth Edition}, Oxford University Press (04, 2021), \href{https://doi.org/10.1093/oso/9780198834625.001.0001}{10.1093/oso/9780198834625.001.0001}.

\bibitem{caracciolo_energy-momentum_1988}
S.~Caracciolo, G.~Curci, P.~Menotti and A.~Pelissetto, \emph{The energy-momentum tensor on the lattice: {The} scalar case}, \href{https://doi.org/10.1016/0550-3213(88)90332-X}{\emph{Nucl. Phys. B} {\bfseries 309} (1988) 612}.

\bibitem{symanzik_continuum_1983}
K.~Symanzik, \emph{Continuum limit and improved action in lattice theories: ({I}). {Principles} and $\phi^4$ theory}, \href{https://doi.org/10.1016/0550-3213(83)90468-6}{\emph{Nucl. Phys. B} {\bfseries 226} (1983) 187}.

\bibitem{balog_puzzle_2010}
J.~Balog, F.~Niedermayer and P.~Weisz, \emph{{The Puzzle of apparent linear lattice artifacts in the 2d non-linear sigma-model and Symanzik's solution}}, \href{https://doi.org/10.1016/j.nuclphysb.2009.09.007}{\emph{Nucl. Phys. B} {\bfseries 824} (2010) 563} [\href{https://arxiv.org/abs/0905.1730}{{\ttfamily 0905.1730}}].

\bibitem{balog_drastic_2012}
J.~Balog, F.~Niedermayer, M.~Pepe, P.~Weisz and U.J.~Wiese, \emph{{Drastic Reduction of Cutoff Effects in 2-d Lattice O(N) Models}}, \href{https://doi.org/10.1007/JHEP11(2012)140}{\emph{JHEP} {\bfseries 11} (2012) 140} [\href{https://arxiv.org/abs/1208.6232}{{\ttfamily 1208.6232}}].

\bibitem{wolff_collective_1989}
U.~Wolff, \emph{Collective monte carlo updating in a high precision study of the x-y model}, \href{https://doi.org/10.1016/0550-3213(89)90236-8}{\emph{Nucl. Phys. B} {\bfseries 322} (1989) 759}.

\bibitem{giusti_energy-momentum_2015}
L.~Giusti and M.~Pepe, \emph{{Energy-momentum tensor on the lattice: Nonperturbative renormalization in Yang-Mills theory}}, \href{https://doi.org/10.1103/PhysRevD.91.114504}{\emph{Phys. Rev. D} {\bfseries 91} (2015) 114504} [\href{https://arxiv.org/abs/1503.07042}{{\ttfamily 1503.07042}}].

\bibitem{giusti_implications_2013}
L.~Giusti and H.B.~Meyer, \emph{{Implications of Poincare symmetry for thermal field theories in finite-volume}}, \href{https://doi.org/10.1007/JHEP01(2013)140}{\emph{JHEP} {\bfseries 01} (2013) 140} [\href{https://arxiv.org/abs/1211.6669}{{\ttfamily 1211.6669}}].

\bibitem{wolff_asymptotic_1990}
U.~Wolff, \emph{Asymptotic freedom and mass generation in the {O}(3) nonlinear $\sigma$-model}, \href{https://doi.org/10.1016/0550-3213(90)90313-3}{\emph{Nucl. Phys. B} {\bfseries 334} (1990) 581}.

\bibitem{makino_renormalizability_2015}
H.~Makino and H.~Suzuki, \emph{{Renormalizability of the gradient flow in the 2D $O(N)$ non-linear sigma model}}, \href{https://doi.org/10.1093/ptep/ptv028}{\emph{PTEP} {\bfseries 2015} (2015) 033B08} [\href{https://arxiv.org/abs/1410.7538}{{\ttfamily 1410.7538}}].

\end{thebibliography}\endgroup

\end{document}